%% file: paper.tex
\documentclass{article}

\newcommand{\subject}{\sloppy \textsf{\textsc{\small{Subject}}}}
\newcommand{\subjects}{\sloppy \textsf{\textsc{\small{Subjects}}}}
\newcommand{\verifier}{\sloppy \textsf{\textsc{\small{Verifier}}}}

\newcommand{\RP}{\textsf{\textsc{\small{RP}}}}
\newcommand{\authenticator}{\sloppy \textsf{\textsc{\small{Authenticator}}}}
\newcommand{\authenticators}{\sloppy \textsf{\textsc{\small{Authenticators}}}}

\newcommand{\authNoutput}{\sloppy \textsf{\textsc{\small{Authenticator Output}}}}
\newcommand{\authNoutputs}{\sloppy \textsf{\textsc{\small{Authenticator Outputs}}}}
\newcommand{\authNmethod}{\sloppy \textsf{\textsc{\small{AuthN Technique}}}}
\newcommand{\authNmethods}{\sloppy \textsf{\textsc{\small{AuthN Techniques}}}}

\newcommand{\authNfactor}{\sloppy \texttt{Authentication Factor}}
\newcommand{\interactionMode}{\sloppy \texttt{Interaction}}
\newcommand{\subjectType}{\sloppy \texttt{Subject}}
\newcommand{\outputMode}{\sloppy \texttt{Output}}
\newcommand{\revocability}{\sloppy \texttt{Revocability}}

\newcommand{\employment}{\sloppy \texttt{Authenticator Employment}}
\newcommand{\sessionTrust}{\sloppy \texttt{Session Trust Contribution}}
\newcommand{\contextSensitivity}{\sloppy \texttt{Contextuality}}
\newcommand{\directionality}{\sloppy \texttt{Directionality}}
\newcommand{\factorType}{\sloppy \texttt{Factor}}
\newcommand{\locality}{\sloppy \texttt{Locality}}
\newcommand{\privacyPreservation}{\sloppy \texttt{Privacy Preservation}}
\newcommand{\uniqueness}{\sloppy \texttt{Uniqueness}}
\newcommand{\subjectInteraction}{\sloppy \texttt{Subject Interaction}}

\newcommand{\tblsubject}{\sloppy \textsf{\textsc{\scriptsize{Subject}}}}
\newcommand{\tblsubjects}{\sloppy \textsf{\textsc{\scriptsize{Subjects}}}}
\newcommand{\tblverifier}{\sloppy \textsf{\textsc{\scriptsize{Verifier}}}}

\newcommand{\tblRP}{\textsf{\textsc{\scriptsize{RP}}}}
\newcommand{\tblauthenticator}{\sloppy \textsf{\textsc{\scriptsize{Authenticator}}}}
\newcommand{\tblauthenticators}{\sloppy \textsf{\textsc{\scriptsize{Authenticators}}}}

\newcommand{\tblauthNoutput}{\sloppy \textsf{\textsc{\scriptsize{Authenticator Output}}}}

\newcommand{\tblauthNmethods}{\sloppy \textsf{\textsc{\scriptsize{AuthN Techniques}}}}

\usepackage{epsfig}
\usepackage{subcaption}
\usepackage{calc}
\usepackage{amssymb}
\usepackage{amstext}
\usepackage{amsmath}
\usepackage{amsthm}
\usepackage{multicol}
\usepackage{pslatex}
\usepackage{algorithm2e}
\usepackage[bottom]{footmisc}
\usepackage[most]{tcolorbox}
\usepackage{enumitem}

\usepackage[T1]{fontenc}
\usepackage{graphicx}
\usepackage{color}
\usepackage{tikz}
\usepackage{listings}
\usepackage{accsupp}
\usepackage{todonotes}
\usepackage{url}
\usepackage{tabularx}
\usepackage{multirow}
\usepackage{longtable}

\usepackage{booktabs}
\usepackage{longtable}
\usepackage{array}
\usepackage{ragged2e}
\usepackage{needspace}

\newcolumntype{L}[1]{>{\raggedright\arraybackslash}p{#1}}
\newcolumntype{Y}{>{\raggedright\arraybackslash}X}

\usepackage{rotating}
\usepackage{graphicx}
\usepackage{placeins}
\usepackage{float}

\graphicspath{{assets/images/}}

\usetikzlibrary{positioning, arrows.meta, shapes.geometric, fit, calc}

\usepackage{arxiv}

\usepackage[utf8]{inputenc} % allow utf-8 input
\usepackage[T1]{fontenc}    % use 8-bit T1 fonts
\usepackage{hyperref}       % hyperlinks
\usepackage{url}            % simple URL typesetting
\usepackage{booktabs}       % professional-quality tables
\usepackage{amsfonts}       % blackboard math symbols
\usepackage{nicefrac}       % compact symbols for 1/2, etc.
\usepackage{microtype}      % microtypography
\usepackage{lipsum}
\usepackage{graphicx}
\usepackage{fancyhdr}
\definecolor{rowgray}{gray}{0.95}

\title{A Faceted Classification of Authenticator-Centric Authentication Techniques}

\author{
  Alex R. Mattukat \\
  Research Group Software Construction\\
  RWTH Aachen University\\
  Ahornstraße 55 \\
  Aachen, Germany \\
  \texttt{mattukat@swc.rwth-aachen.de} 
  \And
  Vincent Schmandt \\
    Research Group Software Construction\\
  RWTH Aachen University\\
  Ahornstraße 55 \\
  Aachen, Germany \\
  \texttt{vincent.schmandt@rwth-aachen.de}
  \And
  Timo Langstrof \\
    Research Group Software Construction\\
RWTH Aachen University\\
  Ahornstraße 55 \\
  Aachen, Germany \\
  \texttt{timo.langstrof@rwth-aachen.de}
  \And
  Michael Zerbe \\
    Research Group Software Construction\\
RWTH Aachen University\\
  Ahornstraße 55 \\
  Aachen, Germany \\
  \texttt{michael.zerbe@rwth-aachen.de}
  \And
  Horst Lichter \\
  Research Group Software Construction\\
  RWTH Aachen University\\
  Ahornstraße 55 \\
  Aachen, Germany \\
  \texttt{lichter@swc.rwth-aachen.de}
  \\
}

\begin{document}
\tiny
This is the \textbf{peer-reviewed, accepted version} of a paper, including minor changes to content that was anonymized for double-blind review. It will appear in the proceedings of the 
21st International Conference on Evaluation of Novel Approaches of Software Engineering (\textbf{ENASE 2026}). The final published version will be available from Science and Technology Publications (SCITEPRESS).
© 2026 SCITEPRESS. Personal use of this material is permitted. Permission from SCITEPRESS must be obtained for all other uses, in any current or future media, including reprinting or republishing this material for advertising or promotional purposes, creating new collective works, resale or redistribution to servers or lists, or reuse of any copyrighted component of this work in other works.

\normalsize
\maketitle
\thispagestyle{fancy}

\begin{abstract}
Authentication is a fundamental security means for protecting system resources. Authenticator-centric authentication techniques (\authNmethods{}) address how mechanisms and credentials are used via \authenticators{}. There are many \authNmethods{} that differ in many ways and there exist classification approaches that aim to structure them. However, they are limited in the aspects they classify and are not flexible enough to accommodate the diverse nature of \authNmethods. This paper presents two contributions. First, novel, faceted classification schemes for \authNmethods{} and \authenticators{} are presented. The schemes were developed based on 345 papers identified through a targeted LLM-assisted literature review and semantic clustering. The classification schemes were applied to build a catalog of \authenticators{} and \authNmethods; the second contribution of this paper. This paper presents our methodology, the classification schemes with example applications, the list of \authNmethods{} from the catalog, and discussions on future work.
\end{abstract}

\keywords{Authentication \and Authenticator \and Authentication Techniques \and Faceted Classification \and Catalog}

\newtcolorbox{mainBox}[1]{%
		enhanced,
		colback=white!90!black,
		colframe=darkgray,
		colbacktitle=darkgray,
		coltitle=white,
		fonttitle=\bfseries,
		title=#1,
		rounded corners,
		boxrule=3pt,
		drop shadow=darkgray,
		width=\textwidth,
	}

\input{content/introduction}

\input{content/foundations}

\input{content/related-work}

\input{content/research-design}

\input{content/classifications}

\input{content/catalog}

% Removed as not being a research artefact
%\input{content/checklist}

\input{content/conclusion}

\bibliographystyle{unsrt}  
\bibliography{Bibliography}

%\cleardoublepage
%\appendix
%\input{Appendix}

\end{document}

%% file: content/introduction.tex
\section{\uppercase{Introduction}} \label{sec:introduction}

Authentication has been around since the early days of computer technology. According to Gersey, MIT's Compatible Time-Sharing System generated the first-ever password in the 1960s \cite{Gersey.2025}. Currently, authentication is a central security measure, as it ensures that only authenticated users, processes, or devices can access restricted system resources \cite{NIST.Authenticator}. This diversity has led to numerous authentication techniques, e.g., usernames and passwords for logging into web applications or certificate-based authentication protocols that rely on public key infrastructures.

Such techniques differ fundamentally in what aspect of authentication they address. They can be divided into two groups: 1) techniques that design how the mechanisms and credentials can be utilized for authenticating, such as the active use of passwords or the passive use of heartbeat patterns; 2) techniques that design how authentication data can be delegated between parties, such as federated authentication. This paper considers only techniques from the first group. We refer to them as ``authenticator-centric authentication techniques'', denoted as \authNmethods{}\footnote{To support the reader, \textsf{\textsc{\small{this font}}} indicates recurring, central terms in this paper.}.

%\subsection{Problem \& Motivation}
The sheer number of \authNmethods{} makes it difficult to select suitable techniques for a given use case, as their suitability depends on many factors. For instance, whether actively entered passwords or passively scanned heartbeat patterns are better suited largely depends on the use case. Consequently, knowing and assessing the strengths and weaknesses of \authNmethods{} is crucial to developing secure software. The complexity of the security domain and the shortage of security experts \cite{Furnell.2020} make this problem even more severe. As a result, non-security experts are often responsible for security-related activities and involved in decision-making \cite{Gutfleisch.2022}. Thus, non-security experts have a particular need for easy access to common knowledge about existing security solutions, such as \authNmethods{}.

\subsection{Goals and Research Questions}
Catalogs of reusable knowledge---such as the well-known Design Pattern Catalog \cite{GoF.1994}---are particularly suited to address these challenges. They provide structured entries that facilitate comparison and navigation. A classification underlying the catalog enables users to filter entries by characteristic properties relevant to their use case. Consequently, a catalog of \authNmethods{} can bridge the gap between security experts, practitioners who need to make informed authentication decisions, and researchers who need to gain an initial understanding of authentication. To our knowledge, no such catalog exists yet. Although existing classifications attempt to categorize authentication approaches by various criteria, they are not compiled into a catalog, which limits their usability and usefulness.

To create a catalog of \authNmethods{}, we need to answer the following two research questions:

\vspace{8pt}
\noindent \textbf{RQ1:} What kinds of \authNmethods{} have been proposed in scientific literature?

\vspace{8pt}

\noindent \textbf{RQ2:} How can \authNmethods{} be classified appropriately based on their characteristic properties?

\vspace{8pt}

\paragraph{Contribution:} 
Since different terms are used in the literature on authentication, we will first define what we mean by an authentication-centric authentication technique (the \authNmethods{}). Next, we show that \authNmethods{} and the \authenticators{} used to generate credentials for authentication are equivalent concepts linked by an aggregation relationship. Based on this, we present a novel classification approach through faceted classification schemes for \authNmethods{} and \authenticators{}. They facilitate the understanding and structuring of these concepts. Finally, we use them to develop a catalog comprising 33 \authNmethods{} and 34 \authenticators{}.

%Next, we propose an online catalog of \authNmethods{}. It contains 33 distinct \authNmethods{} and represents our main contribution. In addition, we present a novel classification approach for \authNmethods{}, which we used to develop the catalog. This classification facilitates the understanding and structuring of the various \authNmethods{}. Overall, our results provide software developers and researchers with a starting point for comparing and selecting different \authNmethods{} based on their characteristic properties.

This paper is structured as follows: Section \ref{sec:foundations} introduces foundations and defines \authNmethods{}. In section \ref{sec:related} we discuss related work. Section \ref{sec:research-design} presents our methodology. The two core concepts and classification schemes are explained in section \ref{sec:classifications}. In section \ref{sec:catalog} we share results from the catalog we developed. Section \ref{sec:conclusion} concludes this paper.

%% file: content/foundations.tex
\section{\uppercase{Foundations}} \label{sec:foundations}

\input{assets/figures/AuthN-process}

To set the scope of the techniques we want to catalog as accurately as possible, a definition of the term \authNmethod{} is needed. It is best defined in terms of the authentication process, which is briefly described below. The process and all presented concepts are based on Todorov \cite{Todorov.2007} and the NIST Digital Identity Guidelines \cite{NIST.Authenticator}.

Figure \ref{fig:AuthN-process} shows the authentication process for successful authentication only, as failed authentication is not relevant for this paper. 

Three entities are involved in this process: \subject{}, \verifier, and \textsf{\textsc{\small{Relying Party (RP)}}}. The \subject{} is the entity whose identity must be verified through authentication; the \verifier{} verifies this identity through authentication. The \RP{} is a system that relies on successful authentication to grant the \subject{} access to its resources or services. 

In addition to these roles, the authentication process is based on the following concepts:
\begin{itemize}
    
    \item \authenticator{}: A mechanism or credential that the \subject{} controls, possesses, or inherently has to prove their identity, such as a password or fingerprint. An \authenticator{} generates \authNoutput{}. 

    \item \authNoutput{}: The data that an \authenticator{} generates. The \verifier{} uses it to confirm that the \subject{} indeed possesses, controls, or inherently has the \authenticator{} that is bound to their claimed identity, or that a condition derived from the environment holds.

\end{itemize}

\noindent The authentication process consists of the following six steps:

\begin{itemize}
    \item[(1)] The \subject{} claims their identity at the \RP{}.
    \item[(2)] The \RP{} requests the \subject{} to authenticate their claimed identity.
    \item[(3)] The \subject{} uses one or more \authenticators{} to validate their identity. They send the created \authNoutput{} to the \verifier{}.
    \item[(4)] The \verifier{} generates the authentication result by verifying the \authNoutput{}.
    \item[(5)] The \verifier{} sends the authentication result to the \RP{} (for \authNmethods{} addressed in this paper, this step is not relevant).
    \item[(6)] In case of a valid authentication result, the \RP{} opens an authentication session for the \subject{}.
\end{itemize}

Step (1) - the identification - is often distinguished from the actual authentication, consisting of (2)-(5) \cite{Todorov.2007}. However, separating them during process realization is not always meaningful. For example, in fingerprint authentication, the fingerprint is used for both identification and actual authentication \cite{fingerprint}. Thus, we do not make this distinction. Furthermore, this process captures the general design of authentication. For instance, in concrete realizations, a \subject{} may send multiple \authNoutputs{} to a \verifier{} before the authentication session is opened.

Based on this process, we define an \authNmethod{} as follows:

\vspace{5pt}

\begin{tcolorbox}[colframe=black, boxrule=0.5pt, arc=0mm, left=2mm, right=2mm, top=1mm, bottom=1mm]
\normalsize
An \textbf{\authNmethod{}} is a conceptual approach to realizing the authentication process. An \authNmethod{} is independent of implementation details, such as authentication data and the protocols used to transmit it.
\end{tcolorbox}

%An \authNmethod{} is a conceptual approach for how a \subject{} uses one or more \authenticators{} to generate \authNoutput{} that enables a \verifier{} to validate the claimed identity, with the goal of accessing resources or services of an \RP. Being conceptual, it is independent of the protocols used to transmit or verify authentication data.

%We can observe that both phases are closely related to each other. In fact, despite being conceptually two distinct phases, there are use cases in which a technical separation of both phases is not meaningful, for instance for reasons of usability. A common example is fingerprint authentication for unlocking hardware devices; when unlocking the device, identification and the actual authentication take place at the same time. %For the sake of completeness, it should be noted that after successful authentication the authorization phase takes place. However, we do not consider it to be part of the authentication process and will therefore not address it further in the remainder of this paper. 

%% file: assets/figures/AuthN-process.tex
\begin{figure}[htb]
    \begin{center}
        \includegraphics[width=0.6\textwidth]{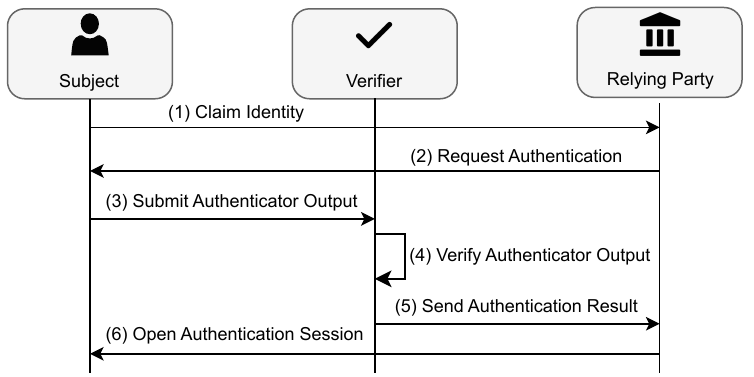}
    \caption{The authentication process for a successful authentication, adopted from NIST \cite{NIST.Authenticator}.}
    \label{fig:AuthN-process}
    \end{center}
\end{figure}

%% file: content/related-work.tex
\section{\uppercase{Related Work}} \label{sec:related}

This work builds on prior work in authentication classification. NIST's Digital Identity Guidelines are a common standard in this field \cite{NIST.Authenticator}. They introduce a categorization of \authenticators{} by authentication factors: \textit{something you know}, \textit{something you have}, and \textit{something you are}. Our approach adopts this categorization. They also define multi-factor \authenticators{} used for Multi-Factor Authentication (MFA). In our work, we present an alternative modeling approach based on aggregation. Moreover, a lot of our terminology and concepts are adapted from the guidelines.

Chenchev et al. extend the classification of NIST by a more fine-grained taxonomy \cite{Chenchev.2021}, based on 1088 reviewed papers. To this end, they divide \authenticators{} of the ``something you are'' category into behavioral and physiological subfactors. We adopt this idea and introduce subfactors for the other factors as well.

Then, some classification approaches aim to differentiate between authentication-related concepts, such as credentials, procedures, and authentication schemes \cite{Alsaeed.2022}. We build on such approaches by designing two separate classification schemes — one for \authenticators{} and one for \authNmethods{} — and by approaching classification through faceted classification schemes.

%Extensive research has focused on classifying \authNmethods{} for specific domains, such as IoT \cite{Mehta.2020} or the medical domain \cite{Alsaeed.2022}. We build on the classification approach proposed by \cite{Alsaeed.2022}. They distinguish between credentials, procedures, schemes, and other aspects of the authentication process. We build on the idea by developing two separate classification schemes — one for \authenticators{} and one for \authNmethods{} — and by approaching classification through faceted schemes.

Lastly, multiple security pattern catalogs have been developed \cite{Schumacher.2006,Fernandez-Buglioni.2013,Berghe.2022}. Such catalogs differ from the catalog we propose in this paper. First, our catalog contains concrete \authNmethods{}, not patterns that describe recurring solutions for implementing \authNmethods{}. Second, most security patterns are at a very high conceptual level, whereas \authNmethods{} operate at the level of specifying which \authenticators{} are used and how, without prescribing implementation details such as authentication data or protocols.

%% file: content/research-design.tex
\section{\uppercase{Research Design}}\label{sec:research-design}

Design science research (DSR) is an established and recognized methodology for developing artifacts and tools. Since the goal of our research is to develop an \authNmethod{} catalog, we followed the six DSR steps as proposed by Peffers et al. \cite{Peffers.2007} as follows:

We presented the identified problem and motivation in section \ref{sec:introduction} (DSR 1). Our research goal is to collect, classify, and catalog \authNmethods{} (DSR 2). We applied an LLM-assisted, targeted literature review and semantic clustering approach to identify \authNmethods{}, created a classification for \authNmethods{}, and used it to create the catalog (DSR 3). The results are presented in an online catalog of \authNmethods{} and in this paper (DSR 4). Given the early stage of the catalog and the fact that similar catalogs improve the application of expert knowledge (e.g., cf. \cite{GoF.1994,Bogner.2019}), we have not yet conducted an evaluation but consider it a central follow-up study (DSR 5). Communication occurs through scientific publications, such as this paper (DSR 6). %Our detailed development approach is presented below.

\subsection{Development Approach} \label{subsec:method}

Our development approach consists of six steps. They are briefly described below.

%\subsubsection{Data Acquisition}
\vspace{8pt}
\noindent \textbf{Step-1 Data Acquisition:} 
Through iterative refinement, we developed a meta-query to identify scientific papers that describe authentication approaches at a conceptual level sufficient for inclusion or exclusion in the catalog according to our definition. %The query is depicted in Listing \ref{lst:SearchQuery}.
We searched the IEEE Xplore database, which yielded 1265 papers. To facilitate data collection, we developed and used a script to download all retrieved papers. Since our goal was to propose a first version of an \authNmethod{} catalog, and given the large number of papers returned by this query, we did not query any other literature databases. Consequently, no duplicates were removed. Nine papers were inaccessible and were therefore excluded, leaving a total of 1256 papers.

%\subsubsection{LLM-assisted Relevance Screening}
\vspace{8pt}
\noindent \textbf{Step-2 LLM-assisted Relevance Screening:}
We filtered relevant papers using an LLM-assisted screening approach based on Tingelhoff et al. \cite{Tingelhoff.2025} and Schulhoff et al. \cite{schulhoffPromptReportSystematic2024}. First, inclusion criteria were defined, requiring papers to introduce authentication approaches with sufficient analytical detail. Moreover, exclusion criteria excluded papers that did not address authentication or that focused on improving, analyzing, evaluating, or applying existing approaches.
We then manually assessed a 5\% random sample $(n=62)$ based on titles and abstracts, establishing ground truth; 19 papers were assessed as relevant, 43 as irrelevant. Using Microsoft's \texttt{MAI-DS-R1} model, we developed a zero-shot chain-of-thought prompt incorporating our definitions and criteria. Through 12 iterations, we refined the prompt and definitions until we achieved 95\% agreement with the ground truth. We aimed for a 95\% agreement, as it corresponds to high inter-rater reliability, allowing the prompt to be used consistently to support relevance assessments across the entire dataset. Finally, we used the latest version of the prompt and definitions with the same \texttt{MAI-DS-R1} model to screen all 1256 papers; 457 were deemed relevant, 799 were excluded.

%\subsubsection{Semantic Clustering}
\vspace{8pt}
\noindent \textbf{Step-3 Semantic Clustering:}
We applied semantic clustering to the resulting dataset using BERTopic to group papers based on the semantic similarity of the \authNmethods{} they address. BERTopic offers a holistic framework for semantic clustering, utilizing Sentence-Bert for semantic embedding, HDBSCAN for clustering, and UMAP for cluster visualization \cite{reimersSentenceBERTSentenceEmbeddings2019,campelloDensityBasedClusteringBased2013,mcinnesUMAPUniformManifold2020}. We used their pre-trained \texttt{all-mpnet-base-v2} model with the paper titles to ensure high performance of the clustering \cite{grootendorstBERTopicNeuralTopic2022,weisserClusteringApproachTopic2020}. We set a minimum cluster size of five to avoid missing potentially relevant small clusters. %; lowering it further produced too many small clusters with no clear underlying topic, which hampered our data analysis. 
25 clusters, comprising 345 of the 457 relevant papers, were identified; 112 papers were identified as outliers.

%\subsubsection{Extracting AuthN Technique Candidates}
\vspace{8pt}
\noindent \textbf{Step-4 Extracting AuthN Technique Candidates:}
Each cluster represented a potential \authNmethod{} candidate. They were extracted in the next step. For each cluster, we selected the highest-cited paper according to IEEE Xplore that met our inclusion criteria. Clusters with no eligible papers were excluded, which was the case for one cluster. Thus, we identified at least one representative paper for 24 clusters, yielding 24 extracted \authNmethods{} candidates. We added five candidates to this list, based on our experience and supported by the literature, that were not included but that we considered \authNmethods{}, yielding a total of 29 \authNmethod{} candidates.

%\subsubsection{Identifying AuthN Techniques} \label{subsec:identification}
\vspace{8pt}
\noindent \textbf{Step-5 Identifying AuthN Techniques:}
We developed an \textit{ontological checklist} to verify whether an \authNmethod{} candidate is a technique according to our definition and can be included in the catalog.  With the second checklist, the \textit{uniqueness checklist}, we ensured that only \authNmethods{} candidates that sufficentily differ from those already included in the catalog were included. The application of these  checklists resulted in one \authNmethod{} candidate being rejected and another being modified. In addition to these 28 \authNmethods{}, five other \authNmethod{} candidates with which we were familiar were positively evaluated, yielding a final set of 33 \authNmethods{}.

%We developed a checklist used to verify whether an \authNmethod{} candidate conforms to our definition of an \authNmethod{} (see section \ref{sec:foundations}). The first author drafted an initial version, which the third and fourth authors independently piloted on five random \authNmethod{} candidates. After discussion to resolve ambiguities, and clarify, supplement, and revise checklist items, the two authors applied the revised checklist to all 29 candidates, achieving 90\% agreement. Cohen's kappa was inapplicable as one author validated all candidates with no negative decisions. Disagreements were resolved through discussion with the first author, leading to another checklist refinement, as well as one exclusion, and one modification of a candidate. Five additional candidates were subsequently proposed based on our experience and validated with the checklist, yielding a final set of 33 \authNmethods{}.

%\subsubsection{Classifying AuthN Technique Candidates} \label{classification-method}
\vspace{8pt}
\noindent \textbf{Step-6 Classifying AuthN Technique:}
For the identified \authNmethods{} to be included in a catalog according to a fixed scheme, they must be classified. To define a suitable classification for \authNmethods{}, we examined the identified \authNmethods{} for characteristic properties, taking existing classifications into account, e.g. \cite{NIST.Authenticator,Chenchev.2021,Alsaeed.2022}. This process yielded two classification schemes, one for \authenticators{}, and one for \authNmethods{}. They are presented in detail in section \ref{sec:classifications}. Both classification schemes were developed through multiple iterations. All identified \authNmethods{} were classified accordingly.

\vspace{8pt}
We have compiled all artifacts from this research into a repository for viewing and analysis \cite{Zenodo}. The repository includes the meta-query used (Step-1), the zero-shot chain-of-thought prompt (Step-2), all paper references resulting from the SLR (Step-3), all identified \authNmethod{} candidates (Step-5), and both checklists along with application examples (Step-6).

%Next, we classified all 33 \authNmethod{} candidates based on recurring properties. To this end, the second author analyzed each candidate and their representative papers, documenting recurring properties. Properties were retained as facets if they applied to all candidates and produced meaningful distinctions between them. For each property, its name, a description of the property, and exemplary candidates were proposed. Then he classified all candidates based on the proposed properties. If a candidate could not be classified by all of them, he refined the properties and re-classified all candidates with the refined properties. Next, the first author reviewed each property and classified the candidates as well. Finally, disagreements were resolved through iterative discussions between the authors until consensus was reached and all candidates were fully classified. 

%% file: content/classifications.tex
\section{\uppercase{Classifying AuthN Techniques}} \label{sec:classifications} 

Analysis of the \authNmethods{} identified revealed that the \authenticators{} that are employed and the way in which they are assembled fundamentally affect the properties of an \authNmethod{} and its suitability in a given use case.  For instance, using a single password for authentication differs significantly from using an MFA approach, which uses both a password and a fingerprint \authenticator{} from both a usability and an architecture perspective. Moreover, usable \authenticators{} for human \subjects{} can be unusable for machine \subjects{}; when designing \authNmethods{} that employ multiple \authenticators{}, this property must thus be taken into account.

It follows that \authenticators{} and \authNmethods{} are core concepts linked by an aggregation relationship. While both core concepts can be characterized by properties that are independent of each other, there are properties of a technique that depend on the properties of the \authenticator{}, too.

Faceted classification schemes can address these requirements. Such classification schemes are considered more flexible and precise than traditional hierarchical ones \cite{Broughton.2006,Priss.2008}. A faceted classification scheme consolidates characteristic properties of the element to be classified into individual, independent facets. Thus, they offer the flexibility needed to model the diverse nature of \authenticators{} and \authNmethods{}.  

Figure \ref{fig:authentication-relationship} depicts the central concepts of our classification approach. \authenticator{} and \authNmethod{} are the core concepts; both are classified by a set of facets that describe their properties. An \authNmethod{} can employ one or more \authenticators{}, indicated by the relationship's cardinality.

\input{assets/figures/authentication-relationships}

%\paragraph{Facet Foundations}
In classification schemes, facets define alternative values of a classification-relevant property. Facets are categorical; they define the alternative facet values on a nominal scale. Facets can be one-dimensional or multi-dimensional. A one-dimensional facet defines mutually exclusive values; i.e., a classified object can take exactly one value at a time, a multi-dimensional facet allows a classified object to take multiple values. Some facets employ hierarchical structures, which is a common facet feature \cite{Priss.2008}.

Moreover, facets can have different levels of importance when classifying objects. A fundamental facet represents a property that plays a fundamental role in the classification of an object. Its value can affect the values of other facets. A common facet is any facet that is not a fundamental facet. It has become common practice to use fundamental facets as the main classes of a classification \cite{Broughton.2006}. In our classification schemes, each \authenticator{} and \authNmethod{} is named after these main classes. 

The faceted classification schemes for \authenticators{} and \authNmethods{} are shown in figures \ref{fig:authenticator-classification} and \ref{fig:technique-classification}. They contain all facets and their facet values that constitute the scheme using the following formatting convention:

\begin{itemize}
    \item[-] Facets are denoted in \texttt{This Format}
    \item[-] \underline{Underlined facets} are fundamental facets
    \item[-] Facets labeled with a * are multi-dimensional
\end{itemize}

In the following, all fundamental and some selected facets of both classification schemes that are particularly important will be explained in detail.

%%%%%%%%%%%%% Authenticator Classification %%%%%%%%%%%%%%%%%%%%%%%%

\input{content/classifications/authenticator}

%%%%%%%%%%%%% AuthN Technique Classification %%%%%%%%%%%%%%%%%%%%%%%%

\input{content/classifications/authN_technique}

\subsection{Naming Convention}
As explained before, it is common practice to name classified items by their fundamental facets. We adhere to this practice and introduce a naming convention for \authenticators{} and \authNmethods{}. It defines a classification name and a readable name. 

The classification name is built by concatenation along the value hierarchies of the facets, and along the two fundamental facets in the case of \authNmethods{}. The value hierarchies are followed from the root to a leaf. We use . as the delimiter between values. Values across facets use | as the delimiter.

\noindent \textit{Example 1:} The classification name of a behavioral \authenticator{} is: 

\texttt{inherence-based.behavioral}.

\noindent \textit{Example 2:} The classification name of an \authNmethod{} that uses a password and a fingerprint in sequential order is:

\texttt{multi.sequential.ordered|multi-factor}

\noindent The readable name is derived from the classification name after removing the delimiters. For brevity, if the \factorType{} value is ``multi-factor'', the ``multi'' value of the \employment{} facet may be omitted in the readable name.

\subsection{Classification Examples}

To conclude, we demonstrate the applicability of our classification schemes by classifying an \authNmethod{} that we refer to as ``Context-aware Touch Authentication'' by \cite{Wang.2019}.  We selected this technique for demonstration purposes because it incorporates all concepts we presented above. It is a \texttt{multi.sequential.ordered|multi-factor} technique, using a \texttt{knowledge-based.free-recall} and an \texttt{inherence-based.behavioral} \authenticator{}. Consequently, the two \authenticators{} employed encompass diverse facets.

Tables \ref{tab:PIN-authenticator} and \ref{tab:touch_authenticator} depict the classified \authenticators{} that are employed by the technique. Table \ref{tab:context-aware} depicts the classified technique.

\input{assets/tables/methods/m1_context-aware}

%% file: assets/figures/authentication-relationships.tex
\begin{figure}[htbp]
    \begin{center}
        \includegraphics[width=0.47\textwidth]{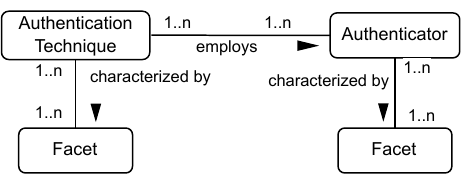}
    \caption{Core classification concepts and their relationships.}
    \label{fig:authentication-relationship}
    \end{center}
\end{figure}

%% file: content/classifications/authenticator.tex
\subsection{Faceted Classification Scheme for Authenticators}

The faceted classification scheme for \authenticators{} is depicted in figure \ref{fig:authenticator-classification}. It defines one fundamental facet, two multi-dimensional facets, and one one-dimensional facet.

The \authNfactor{} facet is fundamental and structured hierarchically by classifying \authenticators{} into exclusive classes that reflect their relationship to the \subject{}. It follows the common approach into inherence-based, knowledge-based, and possession-based \authenticators{}. They build the main classes (factors for short). Factors are further subdivided into finer subfactors, as proposed by Chenchev et al. \cite{Chenchev.2021}. The subfactors categorize \authenticators{} of a factor by a common characteristic. For instance, among possession-based \authenticators{}, some are digital, such as a CA certificate, others are physical, such as a smart card.

Being fundamental, it affects the values of other \authenticator{} facets. For instance, a ``knowledge-based'' \authenticator{} implies that its \interactionMode{} is not passive, as there is no (arguably meaningful) way to provide knowledge passively. Consequently, active interaction is inherent to knowledge-based \authenticators{}.

The \interactionMode{} facet indicates whether an \authenticator{} is suitable for active or passive authentication. A distinction is not always easy to make, as the distinction must be made between the suitability and complete exclusion of an \authenticator{} for active or passive use due to its inherent characteristics. For instance, fingerprints are inherently suited for active authentication, as a \subject{} consciously controls their use. However, they do not exclude passive authentication, for instance via continuous fingerprint scanners attached to the \subject{}'s finger. In contrast, passwords---as a knowledge-based \authenticator{}---cannot meaningfully support passive authentication. Password managers that enter passwords without \subject{} involvement are not comparable. While such tools take over entering a password, conceptually, it must still be entered actively.

%In terms of suitability, a distinction must be made as to whether the \authenticator{}, due to its characteristics, inherently enables active or passive authentication, fundamentally allows a mode, or completely excludes a mode. We illustrate this using fingerprints as an example. In everyday use, they are primarily used for active authentication. This is inherent, as the \subject{} (the human being) has conscious control over their fingers and can therefore actively use them for authentication. However, this does not inherently exclude passive use. For example, an \authNmethod{} that scans the fingerprint in regular intervals with a fingerprint sensor attached to the finger would use the fingerprint for passive authentication. This is fundamentally impossible for a password---a knowledge-based \authenticator{}---as there is no (arguably meaningful) way to query knowledge passively. Password managers that enter passwords without \subject{} involvement are not comparable. While such tools take over entering a password, conceptually, it must still be entered actively. Consequently, the \interactionMode{} of fingerprint \authenticator{} is classified as ``active and passive'', while that of the password \authenticator{} is classified as ``active'' only.

We consolidated the selected facets to the classification scheme because they embody core characteristics of \authenticators{} and thus add value in two respects. First, they can be used to distinguish between suitable and unsuitable \authenticators{} for a given use case. For example, in a medical system where patients in a coma must be authenticated, \authenticators{} that require active control by the \subject{}---such as knowledge-based ones---are unsuitable. Second, these characteristics have architectural implications. A system for active authentication requires a different architecture than one for passive authentication, because the two interaction modes differ in when and how authentication is triggered and performed. Understanding these characteristics can help design suitable architectures and analyze the change impact of adding a certain \authenticator{} to a system.

\input{assets/figures/faceted-classifications/authenticator-classification}

%% file: assets/figures/faceted-classifications/authenticator-classification.tex
\begin{figure}[!t]
    \begin{center}
        \includegraphics[width=0.5\textwidth]{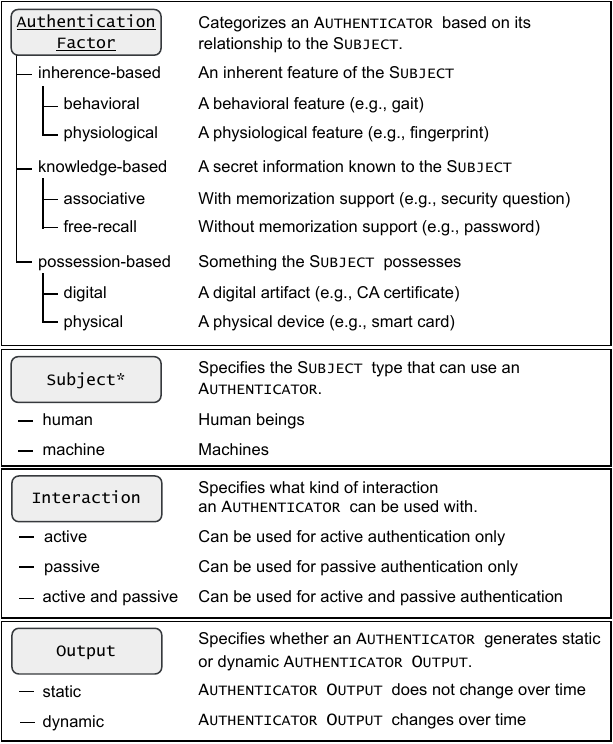}
                \caption{Faceted classification scheme of the \authenticator{}; \textsuperscript{*} indicates a multi-dimensional facet; underlined indicates a fundamental facet.}   

        \label{fig:authenticator-classification}
    \end{center}
\end{figure}

%% file: content/classifications/authN_technique.tex
\subsection{Faceted Classification Scheme for AuthN Techniques}

\input{assets/figures/faceted-classifications/technique-classification}

The faceted classification scheme for \authNmethods{} is depicted in figure \ref{fig:technique-classification}. It defines two fundamental facets, four multi-dimensional facets, and five one-dimensional facets. The consolidated facets were selected for the same reasons as for the \authenticator{} classification scheme.

\employment{} is a fundamental facet. It is structured hierarchically. The first level specifies the cardinality of employed \authenticators{}, i.e., whether the \authNmethod{} employs a single or multiple \authenticators{}. For multiple employments, it specifies whether they are used in parallel or sequentially, and, if used sequentially, whether they must be used in a specific order. Thus, this facet models all possible setups for assembling \authenticators{}.

\factorType{} is the second fundamental facet. It classifies a technique based on the factors that it employs through its \authenticators{}. Subfactors are not included to keep complexity low; if needed, the facet can be modified accordingly in the future. If a technique employs multiple knowledge-based \authenticators{}, it is a knowledge-based \authNmethod{}; if it employs \authenticators{} of different factors, it is a multi-factor technique. %Moreover, \factorType{} and \employment{} have a semantic constraint: \factorType{} can be ``multi-factor'' only if \employment{} is not ``single''.

%This facet also reveals that the \employment{} is the fundamental aspect of this scheme. A technique with a single authenticator cannot be a multi-factor technique, i.e., the value of the \employment{} facet affects the value of the \factorType{}.

Next, we elaborate on \contextSensitivity{}. Some techniques rely solely on the \authenticator{} and its generated \authNoutput{}. Others aim to improve the quality of authentication, for instance, the authentication accuracy, by using additional data that is derived from the context, such as the \subject{}'s environment. %\cite{Al-Rumaim.2024}. 
In our analysis, we identified \authNmethods{} that use spatial, temporal, and state-based data. This facet is optional; if no value is set, a technique is not classified as a contextual one.

Lastly, the \subjectType{} and \subjectInteraction{} facets show a notable, seemingly redundant overlap between the classification schemes. Both are also contained in comparable form in the \authenticator{} classification scheme. However, they are not redundant. The \subject{} types that can use an \authenticator{} and whether it can be used actively or passively are inherent properties of the \authenticator{}. However, it is an inherent property of the \authNmethod{} how authentication is implemented as a whole and which \subject{} types can use it. These properties are determined by the \authenticators{} employed and by the way in which the \authenticators{} are employed. For instance, if an \authNmethod{} uses a password \authenticator{} for active authentication and a fingerprint sensor attached to the \subject{}'s finger for passive authentication, the technique's \subjectInteraction{} as a whole is both ``active'' and ``passive''.

%Overall, the \authNmethod{} classification scheme allows to classify various kinds of employing different \authenticators{} with different properties, to design \authNmethod{} that encompass different properties, that are classifiable by the classification scheme.

%into the classification scheme because they embody core characteristics of \authenticators{} and thus add value in two respects. First, they can be used to distinguish between suitable and unsuitable \authenticators{} per use case. For example, consider a medical system in which human \subjects{} who cannot become active, such as patients in a coma, must be authenticated. For this system, \authenticators{} that must be actively controlled by the \subject{}, such as knowledge-based ones, are unsuitable. Second, these characteristics entail architectural implications. For instance, a system that can actively authenticate \subjects{} requires an architecture different from that of passively authenticating systems. Thus, understanding these characteristics can help design suitable architectures and analyze the change impact of adding a certain \authenticator{} to a system.

%% file: assets/figures/faceted-classifications/technique-classification.tex
\begin{figure}[!t]
    \begin{center}
        \includegraphics[width=0.95\textwidth]{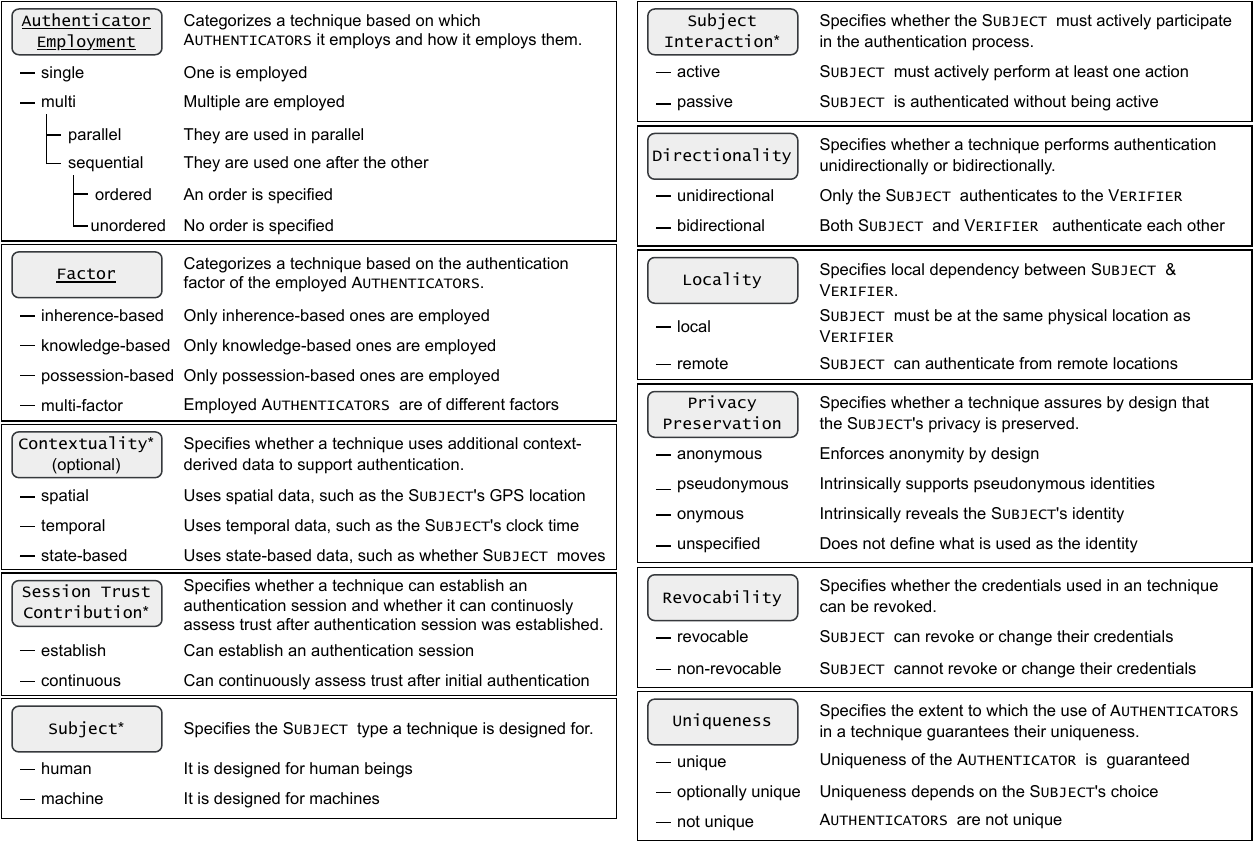}
                        \caption{Faceted classification scheme of the \authNmethod{}; \textsuperscript{*} indicates a multi-dimensional facet; underlined indicates a fundamental facet.}

        \label{fig:technique-classification}

    \end{center}
\end{figure}

%% file: assets/tables/methods/m1_context-aware.tex
%%% Text password authenticator
\begin{table}[!htbp]
\caption{PIN \authenticator{}}
\label{tab:PIN-authenticator}
\scriptsize
\renewcommand{\arraystretch}{1.15}
\rowcolors{3}{rowgray}{white} % start striping from row 3 (optional)

\begin{tabularx}{\linewidth}{L{0.23\linewidth} L{0.22\linewidth} Y}
\toprule
\rowcolor{white}
\textbf{Facet} & \textbf{Value} & \textbf{Reason} \\
\midrule

\authNfactor{} & knowledge-based | free-recall&
PIN is a secret information recalled without memorization support. \\

\interactionMode{} & active &
Knowledge-based \tblauthenticators{} must be provided actively. \\

\subjectType{} & human &
PIN is designed for human \tblsubjects{}. \\

\outputMode{} & static &
The \tblauthNoutput{} of a PIN does not change over time on its own. \\

\bottomrule
\end{tabularx}
\end{table}

%%% Touch interactions Authenticator
\begin{table}[!htbp]
\caption{Touch interactions \authenticator{}}
\label{tab:touch_authenticator}
\scriptsize
\renewcommand{\arraystretch}{1.15}
\rowcolors{3}{rowgray}{white} % start striping from row 3 (optional)

\begin{tabularx}{\linewidth}{L{0.22\linewidth} L{0.23\linewidth} Y}
\toprule
\rowcolor{white}
\textbf{Facet} & \textbf{Value} & \textbf{Reason} \\
\midrule

\authNfactor{} & inherence-based | behavioral&
Touch interactions, i.e., touch rhythm, gesture habits, and keystroke dynamics, are behavioral features. \\

\interactionMode{} & active and passive &
It can be captured passively while typing on the smartphone or used actively. \\

\subjectType{} & human &
Captures unique human touch and movement patterns. \\

\outputMode{} & dynamic &
Behavioral patterns vary over time. \\

\bottomrule
\end{tabularx}
\end{table}

%%% AuthN Technique
\begin{table}[!t]
\caption{Context-Aware Touch \authNmethod{}}
\label{tab:context-aware}
\scriptsize
\renewcommand{\arraystretch}{1.15}
\rowcolors{3}{rowgray}{white} % start striping from row 3 (optional)

\begin{tabularx}{\linewidth}{L{0.25\linewidth} L{0.2\linewidth} Y}
\toprule
\rowcolor{white}
\textbf{Facet} & \textbf{Value} & \textbf{Reason} \\
\midrule

\employment{} & multi.sequential.ordered &
PIN check passes first, then the behavioral layer is invoked. \\

\factorType{} & multi-factor &
Uses a knowledge-based and inherence-based \tblauthenticator{}. \\

\contextSensitivity & state-based &
Body posture (static vs. dynamic) is used. \\

\sessionTrust & establish &
The technique grants or denies access at login only. \\

\subjectType & human &
Both \tblauthenticators{} are designed for humans. \\

\subjectInteraction & active, passive &
PIN is provided actively, touch interactions are provided passively. \\

\directionality & unidirectional &
Only the \tblsubject{} authenticates to the \tblverifier{}. \\

\locality & local &
Authentication runs on-device using embedded phone sensors. \\

\privacyPreservation & onymous &
The \tblRP{} (the smartphone) explicitly identifies a specific registered owner; it intrinsically reveals the \tblsubject{}'s identity. \\

\revocability & non-revocable &
PINs are revocable, touch interactions are not, and the technique does not provide revocability. \\

\uniqueness & unique &
Uniqueness is guaranteed: the phone unambiguously identifies one specific \tblsubject{} among impostors. \\

\bottomrule
\end{tabularx}

\end{table}

%% file: content/catalog.tex
\section{\uppercase{A Catalog of AuthN Techniques}} \label{sec:catalog}

We developed a catalog of \authenticators{} and \authNmethods{} to apply our classification schemes. It is published as an online catalog\footnote{\url{https://scam-research-project.gitlab.io/catalog-of-authentication-techniques/}}. At the time of writing, it contains 34 \authenticators{} and 33 \authNmethods{}. In addition to the facets that classify a catalog entry, the catalog provides a unique name, a description, and a scientific reference that introduces an \authenticator{} or \authNmethod{}. Moreover, \authenticators{} and \authNmethods{} have links to each other, allowing to see details on the \authenticators{} employed by a technique and vice versa. 

We already shared one example from the catalog before: ``Context-aware Touch Authentication''. The complete list of \authNmethods{}, grouped and ordered by their \employment{} and \factorType{} is depicted in table \ref{tab:authN-methods}.

\input{assets/tables/authN_methods}

%% file: assets/tables/authN_methods.tex
\begin{table*}[!htbp]
\caption{The Catalog of \authNmethods{}. Techniques labeled with \textsuperscript{*} are techniques not identified through the literature review. The Technique labeled with \textsuperscript{$\dagger$} is demonstrated in this paper.}
\label{tab:authN-methods}
\scriptsize
\renewcommand{\arraystretch}{1.15}

% ---------- inherence ----------
\rowcolors{3}{rowgray}{white}
\begin{tabular}{
  >{\raggedright\arraybackslash}p{0.573\linewidth}
  >{\raggedright\arraybackslash}p{0.373\linewidth}
}
\toprule
\textbf{single | inherence-based \tblauthNmethods{}  (behavioral)} & \textbf{\tblauthenticator{}} \\
\midrule

Air Handwriting Authentication & Handwriting features \\
Free-text Keystroke Rhythm Authentication & Keystroke rhythm \\
Hand Micro-Movement Authentication & Hand micro-movement patterns \\
Touch Interaction Behavior Authentication &  Touch-interaction behavior \\
Vertical Acceleration Gait Authentication & Vertical foot acceleration patterns \\

\midrule 
\rowcolor{white}
\textbf{single | inherence-based \tblauthNmethods{}  (physiological)} & \textbf{\tblauthenticator{}} \\
\midrule

Face Feature Authentication & Facial feature patterns \\
Fingerprint Authentication\textsuperscript{*}  & Fingerprint features \\
Hand Physiology Authentication & Physiological hand features \\
Heartprint mmWave Radar Authentication & Heartbeat patterns \\
Mobile EEG Authentication & Brain activity patterns \\
PPGPass Wearable Authentication & Photoplethysmogram (PPG) pulse patterns \\
Pulse Active Ratio (PAR) Electrocardiogram (ECG) Authentication & Pulse Active Ratio (PRA) patterns \\
User-Specific Iris Authentication & Iris patterns \\

\bottomrule
\addlinespace
\end{tabular}

% ---------- possession ----------
\rowcolors{3}{rowgray}{white}
\begin{tabular}{
  >{\raggedright\arraybackslash}p{0.573\linewidth}
  >{\raggedright\arraybackslash}p{0.373\linewidth}
}
\toprule
\textbf{single | possession-based \tblauthNmethods{} (digital)} & \textbf{\tblauthenticator{}} \\
\midrule
Anonymous Group Signature Authentication for Vehicular Networks & Group signature certificate \\
Certificate Authentication\textsuperscript{*} & Private cryptographic key with digital certificate \\
One-Time Password Authentication (software-based)\textsuperscript{*} & Initialization key\\
Passkey Authentication\textsuperscript{*} & Private cryptographic key \\

\midrule 
\rowcolor{white}
\textbf{single | possession-based \tblauthNmethods{} (physical)} & \textbf{\tblauthenticator{}} \\
\midrule

Carrier Frequency Offset Authentication & Unique carrier frequency offset characteristics \\
DRAM Physically Unclonable Function (PUF) Authentication & DRAM chip initialization patterns \\
One-Time Password Authentication (hardware-based)\textsuperscript{*} & Hardware OTP Token \\
Smart Card Authentication\textsuperscript{*} & Smart Card \\
Ultralight RFID Authentication & RFID tag \\

\bottomrule
\addlinespace
\end{tabular}

% ---------- knowledge ----------
\rowcolors{3}{rowgray}{white}
\begin{tabular}{
  >{\raggedright\arraybackslash}p{0.573\linewidth}
  >{\raggedright\arraybackslash}p{0.373\linewidth}
}
\toprule
\textbf{single | knowledge-based \tblauthNmethods{}  (associative)} & \textbf{\tblauthenticator{}} \\
\midrule
Cued Graphical Authentication with Timing Intervals & Spatio-temporal graphical password \\
Dynamic Gaze Password (DyGazePass) Authentication & Gaze password \\

\midrule 
\rowcolor{white}
\textbf{single | knowledge-based \tblauthNmethods{} (free-recall)} & \textbf{\tblauthenticator{}} \\
\midrule

PassWalk Authentication\textsuperscript{*} & Textual password  \\
PIN Authentication\textsuperscript{*} & Personal identification number (PIN) \\
Text Password Authentication\textsuperscript{*} & Textual password \\

\bottomrule
\addlinespace
\end{tabular}

% ---------- multiple employed ----------
\rowcolors{3}{rowgray}{white}
\begin{tabular}{
  >{\raggedright\arraybackslash}p{0.37\linewidth}
  >{\raggedright\arraybackslash}p{0.175\linewidth}
  >{\raggedright\arraybackslash}p{0.375\linewidth}
}
\toprule
\textbf{multi.parallel \tblauthNmethods{}} & \textbf{Factor}  & \textbf{\tblauthenticators{}} \\

\midrule

Hand Vein-Knuckle Authentication  & inherence-based & 
Hand vein pattern; Knuckle shape \\

Neuromuscular Password Authentication  & multi-factor &
Textual password; Neuromuscular biometrics 
\\

Multi-Sample Multi-Source Biometric Authentication  & multi-factor  &
Facial features; Voice features
\\
Voice-Teeth Multimodal Authentication & inherence-based  & 
Teeth image; Voice features
\\
Your Song Your Way Rhythm Authentication & inherence-based  & 
Temporal pattern; behavioral features of rhythm input
\\

\midrule 
\rowcolor{white}
\textbf{multi.sequential.ordered \tblauthNmethods{}} & \textbf{Factor}  & \textbf{\tblauthenticators{} (ordered ascendingly)} \\
\midrule

Context-Aware Touch Authentication\textsuperscript{$\dagger$} & multi-factor  & 
PIN; Touch-interaction behavior \\

\bottomrule
\end{tabular}
\end{table*}

%% file: content/conclusion.tex
\section{\uppercase{Conclusion}} \label{sec:conclusion}
This paper presents a novel classification approach for authentication, using faceted classification schemes for authenticator-centric authentication techniques (\authNmethods{}) and the \authenticators{} they employ.  The classification schemes were built based on an LLM-assisted targeted literature review. 24 \authNmethods{} from 1256 initial papers were identified by our review; nine further techniques were subsequently added in the course of this project. 
%Based on these 33 \authNmethods{} in total, a concept model for the two core concepts---\authenticators{} and \authNmethods{}---was developed. Then, for each core concept, a faceted classification scheme was developed through an iterative, targeted analysis of the identified \authNmethods{}. The classification scheme for \authenticators{} consists of four facets, and that for \authNmethods{} of eleven facets. 
Both classification schemes can be used to classify \authNmethods{} and \authenticators{} through a combination of all respective facets. Lastly, both classification schemes were applied by implementing an online catalog of 34 \authenticators{} and 33 \authNmethods{}.

\subsection{Limitations}
Inherence-based techniques are more prominent in the catalog. This becomes particularly apparent for the \authNmethods{} identified in the literature review. 
This is not a critical limitation given our goal of producing an initial catalog, but it stems from several methodological choices: we queried only IEEE Xplore. The database's key areas are, among others, electrical engineering and signal processing. Inherence-based techniques heavily rely on these domains. Additionally, our meta query filtered for novelty indicators, which may favor inherence-based techniques marketed as modern alternatives, while explicitly excluding terms like ``protocol'' and ``cryptographic'' likely filtered out possession-based techniques disproportionately.

The \authenticator{} classification scheme comprises only four facets. In its design, we focused on properties that we could unambiguously consider to be properties of the \authenticator{}. The distinction between properties of it and of an \authNmethods{} is not always easy. For instance, we initially modeled revocability as a property of the authenticator. However, we removed it in a later version, since the semantic meaning of an entity's revocability is unclear and requires greater precision. As this paper represents intermediate results, we consider this limitation not severe.

Next, the classification schemes incorporate solely categorical properties. Thus, it enables only descriptive comparisons. Moreover, it does not allow to model tendencies or weights. For instance, even though fingerprints are far better suited to active authentication, this tendency cannot be modeled in the \contextSensitivity{} facet. In addition, we know that some cross-facet constraints exist, such as that the \factorType{} can only be ``multi-factor'' if the \employment{} is ``multi''. However, such constraints are not yet included. However, we do not consider these critical limitations. This paper presents intermediate results that can serve as a basis for future work to address the limitations. Moreover, the aim of this study was not to achieve qualitative comparability, but rather to develop a systematic classification and description of \authNmethods{}. 

Lastly, the catalog does not represent the state of the art in \authNmethods{}, but the state of research, as the corpus it was built on consists solely of scientific publications.

\subsection{Threats to Validity}
Our results are qualitative in nature. Subjective bias may have influenced them. We attempted to minimize this threat by involving independent researchers and following a rigorous methodology. Each decision was evaluated by at least one other independent researcher, so that the results do not reflect the opinion of a single author. Moreover, any disagreements were discussed bilaterally until a consensus was reached. This applies to any research artefact we developed. %the meta-query we built, the identification of \authNmethods{}, the design of the classification schemes and of the checklists.

Moreover, we used an LLM for abstract screening and clustering; while effective, this approach is relatively new and requires further validation. We validated screening on a random sample; however, results may still be biased by LLM limitations, sampling effects, or prompt overfitting to the validation set, potentially leading to selection bias.

%Literature Review...

\subsection{Future Work}
We plan to extend the catalog by more \authNmethods{} in the future. In particular, the addition of techniques identified in gray literature represents an exciting area for future work. Among other things, this enables comparative studies between techniques that are frequently used in practice and those that are more commonly found in research or niche areas.

In addition, there are multiple opportunities to extend the classification. Additional facets can be designed to model, for instance, the use of \authenticators{} or \authNmethods{} in specific environments, such as IoT, or qualitative properties, such as their usability and accessibility. The schemes can further be extended by embedding cross-facet constraints.

Beyond this, we see further cataloging approaches as important additions to this work. The \authNmethods{} that are listed in our catalog are general approaches to implementing authentication. These are, by design, independent of technical details. A catalog of authentication protocols, i.e., that specify the technical implementation of an \authNmethods{}, would be a valuable addition to our catalog. It would provide a more detailed view of the \authNmethods{}, which is essential for their implementation. Moreover, multiple authentication protocols exist for \authNmethod{}. These could be compared in such a catalog to help developers and architects choose a suitable protocol for implementing an \authNmethod{}. 

Lastly, extending our catalog by authentication patterns and authentication design patterns offers exciting opportunities. While \authNmethods{} are conceptual approaches to realizing the authentication process, security patterns can be used to describe how to realize the techniques. Respectively, authentication design patterns can be formulated to describe how to realize authentication protocols for \authNmethods{}.